
\input phyzzx
\def\e{\adveq\eqno{\rm (\chapterlabel\the\equanumber)}}
\def\adveq{\global\advance\equanumber by 1}

%
%
\rightline{IC/92/301}
\date{Sept 1992}
\titlepage
\vskip 1cm
\title{The Ring Structure of Chiral Operators for Minimal Models Coupled
to 2D Gravity\footnote{*}{To appear in the proceedings of the 1992 ICTP
``Summer School in High Energy Physics and Cosmology".}}
\author { M.H. Sarmadi}
\address{ INFN, Sezione di Trieste \break
and \break
International Centre for Theoretical Physics \break
P.O.Box 586, Trieste, 34100, Italy }
\abstract{ The BRST cohomology ring for $(p,q)$ models coupled to
gravity is discussed. In addition to the generators of the ghost
number zero ring, the existence of a generator of ghost number $-1$
and its inverse is proven and used to construct the entire ring.
Some comments are made regarding the algebra of the vector fields on
the ring and the supersymmetric extension.}
%
%
\def\cmp#1{{\it Comm. Math. Phys.} {\bf #1}}

\def\pl#1{{\it Phys. Lett.} {\bf B#1}}

\def\np#1{{\it Nucl. Phys.} {\bf B#1}}

\def\jmath#1{{\it J. Math. Phys.} {\bf #1}}

\REF\WIT{E. Witten, \np{373} (1992) 187.}
\REF\WZ{E. Witten and B. Zwiebach, preprint IASSNS-HEP-92/4
(January 1992).}
\REF\KMS{D. Kutasov, E. Martinec and N. Seiberg, \pl{276} (1992) 187.}
\REF\F{G. Felder, \np{317} (1989) 215.}
\REF\LZA{B.H. Lian and G.J. Zuckerman, \pl{254} (1991) 417.}
\REF\LZB{B.H. Lian and G.J. Zuckerman, \cmp{145} (1992) 561.}
\REF\BMP{P. Bouwknegt, J.M. McCarthy and K. Pilch, \cmp{145}
(1992) 541.}
\REF\GJJM{S. Govindarajan, T. Jayaraman, V. John and P. Majumdar,
preprint IMSc-91/40, (December 1991).}
\REF\BMPB{P. Bouwknegt, J.M. McCarthy and K. Pilch, preprint
CERN-TH.6279/91, (Oct 1991).}
\REF\IO{K. Itoh and N. Ohta, preprint FERMILAB-PUB-91/228-T.}
\REF\BMPC{P. Bouwknegt, J.M. McCarthy and K. Pilch, \np{377}
(1992) 541.}
\REF\HS{H. Kanno and M.H. Sarmadi, preprint IC/92/150, (July 1992).}
\REF\GJJ{S. Govindarajan, T. Jayaraman and V. John,
preprint IMSc-92/30, (July 1992).}

\vfill \eject
\def\del{\partial}
It was shown by Witten [\WIT] and further elaborated by Witten and Zwiebach
[\WZ] that the physical operators of $c=1$ model (compactified at the
self-dual radius) coupled to 2D gravity have an interesting structure.
The chiral operators of ghost number zero form a ring which is generated
by two elements. Furthermore, the operators of ghost number $+1$ give
rise to vector fields on the ring that satisfy the algebra of area
preserving diffeomorphism. One would like to see if there is an analogous
structure for the minimal models coupled to gravity. There is a
major difference between these models and $c=1$ model. For these models
one has physical states for any ghost number $n_{gh}\in Z$ whereas for $c=1$
case physical states exist only for $-1\leq n_{gh} \leq 2$ (for the chiral
states).

In this talk I will show that the chiral operators of the minimal
models coupled to gravity can be generated by a few operators. The ones
of ghost number zero can be generated [\KMS] by two elements $x$ and $y$ which
are related to the two generators of $c=1$ ring mentioned above by
an $SO(2,C)$ rotation on the two fields of $c=1$, matter and Liouville
$(X,\phi)$. Moreover, there are two operators of ghost numbers $-1$ and
$+1$ denoted by $w$ and $v$, which we will show satisfy $v\cdot w\sim 1$
and therefore $v\sim w^{-1}$, which implies that $w^n \neq 0$ and
$v^n\neq 0$ for all $n\in Z_+$. This means that all the chiral operators
(of the relative cohomology) are of the form $w^nx^iy^j~~~n\in Z,~~
1\leq i\leq p-2,~~ 1\leq j\leq q-2$.

To establish the notation, let us first recall how the chiral operators for
these models are constructed. The most efficient way of obtaining these
operators is by considering the cohomology classes of the BRST operator,
$$
Q_B~=~\oint {dz\over 2\pi i}:\big( T^M(z)+T^L(z)+{1\over 2}T^G(z)
\big)c(z):, \e
$$
where $T^M$ and $T^L$ are the energy-momentum tensors of the matter and
Liouville sectors, with central charges $c^M$ and $c^L=26-c^M$ respectively,
and $T^G $ is the energy-momentum tensor of the ghost system $(b,c)$.
For $(p,q)$ models, in which we will be interested, the matter sector has
the central charge
$$
c_{p,q}~=~1 - {6(q-p)^2\over pq}~,\qquad (p<q,~coprime) \e
$$
and highest weight
$$
\Delta_{r,s}~=~{1\over 4pq} \bigl[ (qr-ps)^2 - (q-p)^2 \bigr]~, \e
$$
$$
(1\leq r \leq p-1,~~1\leq s \leq q-1 )~.
$$

An efficient way of formulating these $(p,q)$ CFT's is through the use of
Coulomb gas. They are described by a single boson $\phi ^M$ which has a
background charge $-\alpha_0$, where $\alpha_0=(q-p)\sqrt{{2\over {pq}}}$.
The space corresponding to the irreducible verma module ${\cal H}_{r,s}$
is then obtained
by using the screening operators $Q_+=\int dz e^{i\alpha_+\phi ^M}$
or $Q_-=\int dz e^{i\alpha_-\phi ^M}$ where $\alpha_+=\sqrt{{{2q}\over p}}$
and $\alpha_-=-\sqrt{{{2p}\over q}}$. It was shown by Felder [\F] that
${\cal H}_{r,s}$ is isomorphic to the cohomology of $Q_F$ where $Q_F$ is
an appropriate power of $Q_+$ or $Q_-$

After coupling these models to Liouville, and describing the Liouville sector
by a free boson with the background charge $-\beta_0=-(\beta_++\beta_-)$
where $\beta_+=\alpha_+$ and $\beta_-=-\alpha_-$, then one obtains the
physical states as the cohomology of $Q_B$ on the complex
${\cal H}_{r,s}\otimes {\cal F}_{\beta}^L\otimes \Lambda ^{b,c}$
where ${\cal F}_{\beta}^L$ is the Fock module whose Liouville charge is
$\beta$ and whose highest weight is $\Delta_{\beta}=
-{1\over 2}\beta(\beta-\beta_0)$.
It is more convenient to restrict oneself first to the relative cohomology
which is the subspace annihilated by $b_0$ and then afterwards
obtain the full space, i.e.,
the absolute cohomology, which has twice as many states as the relative one.

It was proven by Lian and Zuckerman [\LZA,\LZB] and also by Bouwknegt,
McCarthy and Pilch [\BMP] that the relative cohomology
$H^{(n)}({\cal H}_{r,s}\otimes {\cal F}_{\beta}^L\otimes \Lambda ^{b,c},Q_B)$
is non-zero for any integer $n$. As they showed, there is a correspondence
between these states and the singular vectors of Verma module. To see
what these states are, recall the embedding diagram of singular vectors
of the Verma module in the matter sector given below:
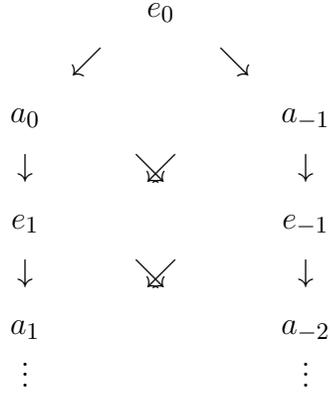
\def\twoarrow{\searrow\kern-11pt\hbox{$\swarrow$}\ }
$$
\matrix{ & & e_0& & \cr
         &\swarrow & & \searrow &  \cr
         a_0& & & & a_{-1}  \cr
         \downarrow& &\twoarrow & & \downarrow  \cr
         e_1& & & & e_{-1}  \cr
         \downarrow& &\twoarrow & & \downarrow  \cr
         a_1 & & & & a_{-2}  \cr
          \vdots& && &\vdots \cr}
$$
In this diagram $a_t$ and $e_t$ are the weights of the singular vectors and are
given by the following expressions:
$$
\eqalign{
a_t &=~ {1\over 4pq} \bigl[ (2pqt + qr + ps)^2 - (q-p)^2
                         \bigr]~,\cr
e_t &=~ {1\over 4pq} \bigl[ (2qpt + qr - ps)^2 - (q-p)^2
                         \bigr]~.\cr}\e
$$
If $\beta$ is chosen such that $1-\Delta _{\beta}$ is equal to one of the
weights in the above diagram then one can find a physical state whose
Liouville charge is $\beta$ and and whose ghost number is $n_{gh}=d_{\beta}
sign(\beta-{1\over 2}\beta_0)$ where $d_{\beta}$ is the number of steps from
that particular node to top node ($e_0=\Delta_{r,s}$). Solving the equation
$1-\Delta_\beta~\in~\{ a_t,e_t\}$ for $\beta$ one finds
that for weights of type $a_t$,
$$
\beta={1\over \sqrt{2pq}}\bigl(p+q\pm(2pqt+qr+ps)\bigr),  \e
$$
and for those of type $e_t$,
$$
\beta={1\over \sqrt{2pq}}\bigl(p+q\pm(2pqt+qr-ps)\bigr).  \e
$$
Using the notation $\beta_{r,s}={{1-r}\over 2}\beta_++{{1-s}\over 2}\beta_-$
where $\beta_+=\sqrt{2q/p}$ and $\beta_-=\sqrt{2p/q}$, from the above
expressions for the Liouville charges we see that in the $(r,s)$ sector
the states of ghost number $-1$ (which give rise to operators of ghost number
zero) have the charges $\beta _{r,s}$ and $\beta _{p-r,q-s}$. In the sectors
$(1,2)$ and $(2,1)$ one can easily write two such ghost number zero operators:
$$
x=\bigl( bc+(i\alpha_{1,2}\del \phi^M+\beta_{1,2}\del \phi^L)\bigr)
       e^{i\alpha_{2,1} \phi^M+\beta_{2,1} \phi^L}~, \e
$$
and
$$
y=\bigl( bc+(i\alpha_{2,1}\del \phi^M+\beta_{2,1}\del \phi^L)\bigr)
       e^{i\alpha_{1,2} \phi^M+\beta_{1,2} \phi^L}~.  \e
$$
In fact these two operators are just the two generators of $c=1$ chiral
ring after an $SO(2,C)$ rotation on the (matter, Liouville) field space.
Then the two ghost number zero operators in the $(r,s)$ sector whose
Liouville charges are $\beta_{r,s}$ and $\beta_{p-r,q-s}$ are respectively
$x^{r-1}y^{s-1}$ and $x^{p-r-1}y^{q-s-1}$.

Further inspection of eqs. (5,6) for ghost numbers $-2$ and $+1$ states
(i.e. ghost numbers $-1$ and $+1$ operators) shows that in the sector
$(p-1,1)=(1,q-1)$ there are two operators, which we denoted as $w$ and $v$,
which have Liouville charges $\beta_w=\beta_{1,q+1}$ and $\beta_v=-\beta_w$.
This suggests that if $v\cdot w$ is non-zero then $v\cdot w\sim 1$. Moreover,
from eqs. (5,6) one sees that all the Liouville charges are of the form
$n\beta_w + \beta_{m,m'}$ for $n\in Z~,~1\leq m\leq p-1$ and $1\leq m'\leq
q-1$.
Therefore, if $v\cdot w$ does not vanish then any power of $w$ and $v=w^{-1}$
would be non-zero and one should be able to obtain ghost number $-n$
operators by taking the product of the $n$'th power of the operator $w$
and the ghost number zero operators, i.e., all the operators are of the form
$w^nx^{m-1}y^{m'-1}$. Before proving the non-vanishing of $v\cdot w$
for the general $(p,q)$ models, let us look at
two examples, namely the models $c^M=0$ and $c^M={1\over 2}$.

\noindent {\it {(2,3) Model}}\hfill \break
For this example there are two operators of ghost number zero, namely
the identity operator and the operator $y$. One can also explicitly find
the operator $w$ of ghost number $-1$ and Liouville charge
$\beta_{3,1}=-\beta_+=-\sqrt{3}$ by solving the equation
$$
[Q_B,w]=0~ {\rm mod ~null ~operators}.  \e
$$
One finds the follwing expression for $w$:
$$
w~=~\big( b\del b c~-~{1\over\sqrt3} b\del^2 \phi^L
       ~+~{1\over 2\sqrt3} \del b\del\phi^L ~+~{1\over6}\del^2 b
        \big)~e^{-\sqrt3\phi^L}~. \e
$$
The operator $v$ of ghost number $+1$ and Liouville charge $\beta_+$ can be
represented by \break $ce^{i\alpha_0\phi^M+\beta_+\phi^L}$ or by
$ce^{\beta_+\phi^L}$ since both $1$ and $e^{i\alpha_0\phi^M}$ represent
the identity in the coulomb gas description of the matter sector. In any case
after working out the operator product expansion one finds
$$
v(z)w(0)=-{1\over 6}{\bf 1},
$$
and therefore the relative cohomology is generated by $w$ and $y$. In general,
as in the case of $c=1$ of ref. [\WZ], to obtain the absolute cohomology
which has twice as many states as the relative cohomology, one
multiplies the operators in the relative cohomology by the physical
operator
$$
a~=~c\del \phi+{1\over 2} \beta _0\del c ~. \e
$$
Thus all the operators are of the form $w^ny^ia^k$ where $n\in Z$ and
$i,k=0,1$.

Two comments are in order. First, one can write a current whose charge
acts on the ring as the vector field $w\partial _a$. This current should have
ghost number $-2$ and the action of $Q_B$ on it should give a total derivative.
For this example, the only such current one can write is $b\partial b
e^{-\sqrt{3}\phi^L}$ and it is in fact $(b_{-1}w)=\oint b(z)w$. and therefore
$$
[Q_B,\oint b\partial b e^{-\sqrt{3}\phi^L}]=0.
$$
One can explicitly check that it acts on the ring as $w\partial _a$. The second
comment concerns eq. (9) which determines $w$. One has the freedom of
adding null operators to $w$ and thereby change the  representative for
$w$. In fact by adding the term ${1\over 4}({i\over \sqrt{3}}\partial^2\phi ^M
-(\partial\phi ^M)^2)e^{-\sqrt{3}\phi ^L}$
to $w$ given in eq. (10) one obtains a representative which satisfies the
equation $[Q_B,w]=[Q_+,x^2]$. In general one can choose a representative
for $w$ which satisfies
$$
[Q_B,w]=[Q_+,x^p].
$$

\noindent {\it (3,4) model}\hfill \break
For this example, the ghost number zero operators are
$x^iy^j~~i=0,1,~~j=0,1,2$.
To show that $v\cdot w$ is non-zero it is simpler to write first $w\partial _a$
and then show that its action on $av$ is not zero. To write $w\partial _a$
one needs to write a current of ghost number $-2$ on which the action of
$Q_B$ gives a total derivative up to null operators. One finds that
it is given by the following expression
$$
\eqalign{
j^{(-2)}(z)=&\biggl[\bigl[(-{5\over 2\sqrt{6} } \partial^2\phi^L
              -{1\over 2}(\partial\phi^L)^2)
              -{11\over 9}(-{3\over 2\sqrt{6}}~i\partial^2\phi^M
                  -{1\over 2}(\partial \phi^M)^2) \bigr]b\partial b \cr
              &-{4\over 3\sqrt{6}}~i\partial \phi^M b\partial^2 b
            +{1\over 3} b\partial^3 b-{11\over 12}\partial b\partial^2b\biggr]
             e^{-2/\sqrt{6}~i\phi^M-\sqrt{6}\phi^L}. \cr} \e
$$
Using this expression one finds that the action of $\oint j^{(-2)}$ on
$av=a(z)c(0)e^{3i/\sqrt{6}\phi^M
+\sqrt{6}\phi^L}=-{5\over 2\sqrt{6}}~\partial cc
e^{3i/\sqrt{6}\phi^M+\sqrt{6}\phi^L} $ is non-zero.

Now we consider the general case of $(p,q)$ models. In order to prove
that the product $v\cdot w$ does not vanish, we will show that a
correlator of the type $<w~v~{\cal O}^{(3)}>$ on the sphere is non-zero.
Here the operator ${\cal O}^{(3)}$ is a physical operator of ghost number $3$
and Liouville charge $\beta_0$. It is just the product of the operator $a$
with the operator $\del ^2cce^{\beta_0\phi}$.
This latter operator satisfies the following equation [\LZB,\GJJM]:
$$
\del ^2cce^{\beta_0\phi}=[Q_B~,~c\del \phi^Me^{\beta_0\phi^L}]. \e
$$
Note that the operator $c\del \phi^Me^{\beta_0\phi^L}$ is not a physical
operator since $\partial \phi^M$ does not correspond to any of the states
in the irreducible module. now inserting the right hand side of eq. (13)
inside the correlator and pulling the contour of $Q_B$ to have it act on $w$
we can then use
$$
[Q_B~,~w]=[Q_+~,~x^p]. \e
$$
Now the contour of $Q_+$ can be pulled to have it act on
$c\partial \phi^M e^{\beta_0 \phi^L}$ to give $ce^{i\alpha_{-1,1}\phi^M
+\beta_0\phi^L}$. One can also show that
$$
x^p~v~\sim ~ce^{i\alpha_{1,-1}\phi^M}. \e
$$
Therefore, the above correlator is reduced to the following two
point function
$$
\langle \alpha_{1,-1},0|~c_{-1}c_0c_1~|\alpha_{-1,1},\beta_0 \rangle ,
$$
which is clearly non-zero. This implies that the product $v\cdot w$ cannot
vanish and we have therefore proven the proposed ring structure.
Note that this ring is a non-commutative one, the operators $x,~y$ and $a$
commute with each other but anti-commute with $w$.

For the above ring structure, since all integer powers of $w$ are present,
it is natural to expect a Virasoro algebra for the vector fields.
One can construct the vector
fields as follows. Given a physical operator $\psi^{(n)}$ of ghost number $n$
one can construct $\oint (b_{-1}\psi^{(n)})$ which has ghost number $n-1$
and commutes with $Q$. It acts as a vector field on the ring. By this
construction one finds that for the operators of type $w^nx^iy^j$ the
vector fields are
$$
G^{i,j}_n=w^nx^iy^j\partial_a ~,
$$
and for those of the type $aw^nx^iy^j$ the vector fields are
$$
K^{i,j}_n=-x^iy^jw^n\big(w\del _w+{1\over p}x\del _x+{1\over q}y\del _y
          -(n+{i\over p}+{j\over q})a\del _a\big)~. \e
$$
After introducing the appropriate cocycle factors one finds that
they satisfy the following algebra:
$$
\eqalign{
[{\tilde K}^{i,j}_n,{\tilde K}^{k,l}_m]&=(n-m+{{i-k}\over p}
+{{j-l}\over q}){\tilde K}^{i+k,j+l}_{n+m}~, \cr
[{\tilde K}^{i,j}_n,{\tilde G}^{k,l}_m]&=-(n+m+{{i+k}\over p}
+{{j+l}\over q}){\tilde G}^{i+k,j+l}_{n+m}~, \cr
[{\tilde G}^{i,j}_n,{\tilde G}^{k,l}_m]&=0 ~ . \cr }
$$
Therefore the vector fields ${\tilde K}^{0,0}_n$ satisfy a Virasoro algebra
under which ${\tilde K}^{i,j}$ and ${\tilde G}^{i,j}$ are primaries
of weights $2$ and $0$ respectively.

One can also consider the ring structure for $N=1$ super-minimal
models. Assuming that one can generalize Felder's discussion to these models,
in refs. [\BMPB,\IO,\LZB] it was pointed out that the results about the
BRST chomology carry over to the supersymmetric case, that the cohomology is
non-trivial for all ghost numbers. For ${\hat c}=1$, it has been shown
[\BMPC] that the ring of ghost number zero operators is also generated
by two elements. After the appropriate $SO(2,C)$ rotation one obtains
the ghost number zero ring for the super-minimal models. Moreover, one can
write the Liouville charges for all the non-zero ghost number operators by
solving the equation ${1\over 2}-\Delta_\beta~\in~\{a_t,e_t\}$, where
$\{a_t,e_t\}$ are the set of weights of singular vectors for the super
conformal case. One finds that again all these charges are of the form
$n\beta_w+\beta_{m,m'}$ which suggests that the non-zero ghost number
operators are obtained from integer powers of an element $w$.

The talk presented here was based on the work [\HS] done in
collaboration with H. Kanno whom I would like to thank. A different
ring structure for the minimal models has been proposed in ref. [\GJJ].

\refout

 \end